\documentclass[10pt,conference]{IEEEtran}
\usepackage{cite}
\usepackage{amsmath,amssymb,amsfonts}
\usepackage{algorithmic}
\usepackage{graphicx}
\usepackage{textcomp}
\usepackage{xcolor}
\usepackage[hyphens]{url}
\usepackage{fancyhdr}
\usepackage{hyperref}

\usepackage{pifont}
\usepackage{amssymb}
\usepackage{cellspace}

\usepackage{tikz}   
\usepackage{xcolor}
\newcommand*\circled[1]{\tikz[baseline=(char.base)]{
            \node[shape=circle,fill,inner sep=1pt] (char) {\textcolor{white}{#1}};}}

\newcommand{\hpcayear}{2025}
\newcommand{\cmark}{\ding{51}}%
\newcommand{\xmark}{\ding{55}}%

\definecolor{orange}{RGB}{240, 140, 0}
\definecolor{green}{RGB}{0, 158, 0}
\definecolor{blue}{RGB}{0, 0, 210}

\usepackage{tablefootnote}
\usepackage{threeparttable}

\newcommand{\hpcasubmissionnumber}{1441}
\title{FIGLUT: An Energy-Efficient Accelerator Design for \underline{F}P-\underline{I}NT \underline{G}EMM Using \underline{L}ook-\underline{U}p \underline{T}ables}

\def\hpcacameraready{} 

\newcommand\hpcaauthors{Gunho Park$^{1,2*}$, Hyeokjun Kwon$^{1*}$, Jiwoo Kim$^{1}$, Jeongin Bae$^{2}$, Baeseong Park$^{2}$, \\Dongsoo Lee$^{2}$, and Youngjoo Lee$^{1}$}
\newcommand\hpcaaffiliation{$^{1}$Pohang University of Science and Technology (POSTECH), \\
$^{2}$NAVER Cloud}
\newcommand\hpcaemail{\textit{\{gunho.park3\}@navercorp.com, \{kwon36hj, youngjoo.lee\}@postech.ac.kr}}

\author{
  \ifdefined\hpcacameraready
    \IEEEauthorblockN{\hpcaauthors{}}
      \IEEEauthorblockA{
        \hpcaaffiliation{} \\
        \hpcaemail{}
      }
  \else
    \IEEEauthorblockN{\normalsize{HPCA \hpcayear{} Submission
      \textbf{\#\hpcasubmissionnumber{}}} \\
      \IEEEauthorblockA{
        Confidential Draft \\
        Do NOT Distribute!!
      }
    }
  \fi 
}

\fancypagestyle{camerareadyfirstpage}{%
  \fancyhead{}
  
  \fancyhead[C]{
    \ifdefined\aeopen
    \parbox[][12mm][t]{13.5cm}{\hpcayear{} IEEE International Symposium on High-Performance Computer Architecture (HPCA)}    
    \else
      \ifdefined\aereviewed
      \parbox[][12mm][t]{13.5cm}{\hpcayear{} IEEE International Symposium on High-Performance Computer Architecture (HPCA)}
      \else
      \ifdefined\aereproduced
      \parbox[][12mm][t]{13.5cm}{\hpcayear{} IEEE International Symposium on High-Performance Computer Architecture (HPCA)}
      \else
      \parbox[][0mm][t]{13.5cm}{\hpcayear{} IEEE International Symposium on High-Performance Computer Architecture (HPCA)}
    \fi 
    \fi 
    \fi 
    \ifdefined\aeopen 
      \includegraphics[width=12mm,height=12mm]{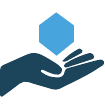}
    \fi 
    \ifdefined\aereviewed
      \includegraphics[width=12mm,height=12mm]{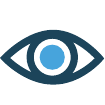}
    \fi 
    \ifdefined\aereproduced
      \includegraphics[width=12mm,height=12mm]{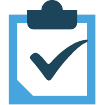}
    \fi
  }
  \fancyfoot[C]{}
}
\begin{document}
\maketitle

\ifdefined\hpcacameraready 
  \thispagestyle{camerareadyfirstpage}
  \pagestyle{empty}
\else
  \thispagestyle{plain}
  \pagestyle{plain}
\fi

\newcommand{\hpcaheight}{0mm}
\ifdefined\eaopen
\renewcommand{\hpcaheight}{12mm}
\fi

\def\thefootnote{*}\footnotetext{Equal contribution.}

\begin{abstract}

Weight-only quantization has emerged as a promising solution to the deployment challenges of large language models (LLMs).
However, it necessitates FP-INT operations, which make implementation on general-purpose hardware like GPUs difficult.
In this paper, we propose FIGLUT, an efficient look-up table (LUT)-based GEMM accelerator architecture. 
Instead of performing traditional arithmetic operations, FIGLUT retrieves precomputed values from an LUT based on weight patterns, significantly reducing the computational complexity.
We also introduce a novel LUT design that addresses the limitations of conventional memory architectures.
To further improve LUT-based operations, we propose a half-size LUT combined with a dedicated decoding and multiplexing unit.
FIGLUT efficiently supports different bit precisions and quantization methods using a single fixed hardware configuration.
For the same 3-bit weight precision, FIGLUT demonstrates 59\% higher TOPS/W and 20\% lower perplexity than state-of-the-art accelerator design.
When targeting the same perplexity, FIGLUT achieves 98\% higher TOPS/W by performing 2.4-bit operations.

\end{abstract}

\section{Introduction}
Pretrained Large Language Models (LLMs) have demonstrated remarkable performance in a wide range of language understanding and generation tasks.
These models, predominantly based on the Transformer architecture \cite{vaswani2017attention, sun2023retentive, anil2023palm, genc2021gemmini, achiam2023gpt4}, exhibit a predictable power-law scaling in performance relative to their parameter size.
Consequently, the size of the model has increased to unprecedented levels, pushing the boundaries of what can be achieved in natural language processing.
However, substantial growth in model size comes with significant challenges, particularly in terms of memory footprint. \cite{luccioni2023estimating, yoo2023tf}
For instance, a model with 175 billion parameters, requiring approximately 350GB of memory, far exceeds the DRAM size of 80GB available in current state-of-the-art GPUs \cite{park2022lut}.
Additionally, memory bandwidth also becomes a critical issue, leading to significant memory bottlenecks.
This bottleneck can severely impact the performance and efficiency of deploying such large models, as data transfer rates cannot keep up with the computational demands of the LLMs.
This discrepancy highlights the urgent need for efficient methods to deploy such large models on hardware accelerator.

One of the most promising approaches to mitigate the memory-related challenges of LLM is parameter quantization~\cite{wang2023bitnet, chee2024quip, huang2024billm}.
In the context of LLMs, there has been a shift toward weight-only quantization, where only weights are strongly quantized to sub-4-bit precision while activations remain in floating point (FP) format \cite{bai2018proxquant, chung2020extremely}.
This shift is motivated by the observation that the memory footprint of the weights significantly exceeds that of the activations in the LLM, and the presence of outliers in the activations further supports the effectiveness of weight-only quantization. 
Despite its advantages, weight-only quantization requires the use of non-standard FP-INT operations, posing an ongoing challenge.

Recent studies have introduced efficient kernels for weight-only quantization, enabling practical speed-ups on GPUs \cite{frantar2022optq, lin2024awq, park2022lut, you2024shiftaddllm}.
Although the kernels proposed in \cite{frantar2022optq, lin2024awq} demonstrated faster speeds compared to cuBLAS, which is a GPU-accelerated library to accelerate HPC applications, including GEMM operation, these improvements largely stem from efficient data movement due to weight compression. 
However, during computation, the compressed weights are dequantized back to FP format, resulting in inefficiencies as the actual arithmetic operations are still performed using FP-FP units.
NVIDIA's recent release of new FP-INT GEMM kernels in CUTLASS \cite{kim2022says} similarly relies on inherent FP operations.
Efforts to overcome these inefficiencies have led to the proposal of LUT-based GEMM kernels \cite{park2022lut, you2024shiftaddllm}, which aim to perform FP-INT GEMM operations without dequantization. 
However, storing LUTs in shared memory on GPUs often leads to bank conflicts, causing inefficiencies because multiple threads access the same shared memory bank simultaneously.

To address these issues, recent advancements propose novel hardware architectures designed to efficiently perform FP-INT operations specifically tailored for weight-only quantized models \cite{kim2023squeezellm}. These emerging solutions aim to bridge the gap between the theoretical benefits of weight-only quantization and practical deployment on existing hardware.
For instance, iFPU \cite{kim2023winning} proposes an efficient bit-serial architecture accelerator utilizing binary-coding quantization (BCQ) format.
In the BCQ format, the inner product of binary weights and input activations is replaced by the sum or subtraction of input activations.
iFPU introduces a pre-alignment technique that aligns the exponents of input activations with FP values.
Consequently, the operations on pre-aligned mantissas are handled by integer arithmetic units, efficiently performing FP-INT operations.
Similarly, FIGNA \cite{jang2024figna} presents FP-INT4 multiplication employing the pre-alignment technique while addressing the inefficiencies inherent in the bit-serial structure adopted by iFPU.
By using pre-alignment, FIGNA replaces FP-INT multiplication with integer multiplication between aligned mantissas and quantized weights, enhancing the efficiency of weight-only quantized model computations.
However, FIGNA has limitations as it is fixed to a specific precision (e.g., 4-bit quantized weights), making it inefficient for models quantized to less than 4-bit precision or for those using mixed-precision quantization methods.

In this paper, we propose FIGLUT, a novel accelerator architecture that leverages LUT-based GEMM to reduce the computational complexity of bit-serial accelerators.
FIGLUT replaces arithmetic operations with table look-ups using weight patterns as keys, significantly enhancing computational efficiency.
We introduce a specialized operator, the read-accumulate (RAC) unit, which replaces the conventional multiply-accumulate (MAC) unit found in traditional hardware accelerators.
FIGLUT efficiently supports FP-INT operations for weight-only quantized models and employs a bit-serial structure to manage computations of various precisions within a single hardware framework.
Moreover, by using the BCQ format for weight representation, FIGLUT can operate not only on BCQ models, which present state-of-the-art performance, but also on commonly used uniformly quantized models.
As a result, FIGLUT enables efficient computation across different bit precisions and quantization methods on a single hardware platform, representing a significant advancement in the deployment of weight-only quantized models.

Our major contributions in this work include the following:
\begin{itemize}
    \item We propose an LUT-based FP-INT GEMM method that replaces arithmetic operations with LUT reads, facilitating energy-efficient GEMM operations. 
    \item We introduce a novel specialized LUT architecture that enables simultaneous access to the LUT without bank conflicts during parallel processing.
    \item We present FIGLUT, an innovative LUT-based FP-INT GEMM accelerator that efficiently supports various quantization methods and precisions on a single hardware platform by leveraging LUT-based operations.
\end{itemize}

\section{Background}

\subsection{Methods and Systems for Weight-Only Quantization}

\renewcommand{\arraystretch}{1.3}
\begin{table}[]
\centering
\caption{Comparison of different hardware accelerators}
\label{tab:comparison_accelerators}
\begin{tabular}{ccccc}
\hline
                           & \begin{tabular}[c]{@{}c@{}}FP-INT\\ Operation\end{tabular} & \begin{tabular}[c]{@{}c@{}}Mixed-\\ Precision\end{tabular} & \begin{tabular}[c]{@{}c@{}}BCQ\\ Support\end{tabular} & \begin{tabular}[c]{@{}c@{}}Computational\\ Complexity\end{tabular} \\ \hline
GPU                        & \xmark                                                     & \xmark                                                    & \xmark                                                        & $\mathcal{O}(mnk)$                                                 \\
iFPU \cite{kim2023winning} & \cmark                                                     & \cmark                                                    & \cmark                                                        & $\mathcal{O}(mnkq)$                                                \\
FIGNA \cite{jang2024figna} & \cmark                                                     & \xmark                                                    & \xmark                                                        & $\mathcal{O}(mnk)$                                                 \\
Proposed                   & \cmark                                                     & \cmark                                                    & \cmark                                                        & $\mathcal{O}(mnkq/\mu)$                                            \\ \hline
\end{tabular}
\end{table}

Quantization is a powerful method for reducing computational complexity and memory footprint by converting high-bit precision FP values into low-bit INT values.
However, applying quantization to LLMs presents significant challenges, especially when attempting to quantize both weights and activations to lower bit precisions.
Activation values, in particular, are more difficult to quantize than weight values due to the presence of outliers and their dynamic nature, which makes their distribution hard to predict accurately \cite{yao2022zeroquant, dettmers2022gpt3}.
This complexity necessitates specialized techniques to effectively manage the quantization of activations in LLMs.

To address memory-related challenges such as memory bound of LLM inference and limited DRAM size, weight-only quantization methods have been extensively studied.
These methods preserve accuracy by keeping the activations in FP format.
Due to the significant accuracy drop observed with simple round-to-nearest (RTN) quantization, various techniques have been proposed to minimize quantization error.
OPTQ \cite{frantar2022optq} introduces a uniform quantization method that utilizes approximate second-order information from a calibration dataset to achieve sub-4-bit quantization with negligible accuracy loss.
AWQ \cite{lin2024awq} proposes a weight-only quantization approach that selects salient weights based on activation distribution to minimize quantization error.
Recently, non-uniform quantization techniques using BCQ have also demonstrated excellent performance.
ShiftAddLLM \cite{you2024shiftaddllm} achieves state-of-the-art results by applying post-training bitwise shift-and-add reparameterization with non-uniform BCQ, enabling a multiplication-less LLM.
Additionally, ShiftAddLLM improves the accuracy-efficiency trade-off by employing mixed-precision quantization based on the sensitivity of each layer.

With the proven efficiency of weight-only quantization methods, various hardware accelerators have been proposed to support FP-INT operations between activations and weights.
Table~\ref{tab:comparison_accelerators} compares the features of different hardware accelerators. 
Commercial GPUs lack FP-INT arithmetic units that can handle FP input activations and INT weights.
As a result, a dequantization process is necessary to convert INT weights back to FP format for use with existing FP-FP arithmetic units.
The FP-INT GEMM kernels proposed in \cite{frantar2022optq} and \cite{lin2024awq} achieve better latency compared to cuBLAS FP-FP GEMM, thanks to efficient bandwidth utilization enabled by weight compression in memory-bound LLM applications.
However, the absence of native FP-INT operations on GPUs undermines the advantages of weight quantization by necessitating FP-FP computations.

To address inefficiencies in GPUs and effectively handle FP-INT GEMM operations, several accelerator designs have been proposed.
iFPU \cite{kim2023winning} introduces a method to efficiently process computation between weights in BCQ format and FP activations by pre-aligning the mantissas of input activations based on the maximum exponent value.
This alignment allows FP-INT operations to be replaced by INT-INT addition between mantissas and binary weights, reducing hardware complexity. 
While this bit-serial approach enables support for mixed-precision quantization, it inherently increases computational complexity in proportion to the bit precision $q$ of the quantized weights.

FIGNA \cite{jang2024figna} addresses the computational overhead of bit-serial architecture by proposing an FP-INT arithmetic unit that performs INT-INT multiplications between pre-aligned mantissa and uniformly quantized weights.
While this approach effectively reduces computational complexity, it is constrained by its fixed precision, limiting its applicability to uniformly quantized models and precluding support for sub-4-bit precision or advanced BCQ-based non-uniform quantization methods.
Consequently, there is a need for a new bit-serial hardware architecture that minimizes computational overhead while supporting a broader range of precisions and quantization methods, including non-uniform quantization with BCQ.

\subsection{Binary Coding Quantization}

\begin{figure}[]
  \centering
  \includegraphics[width=0.90\linewidth]{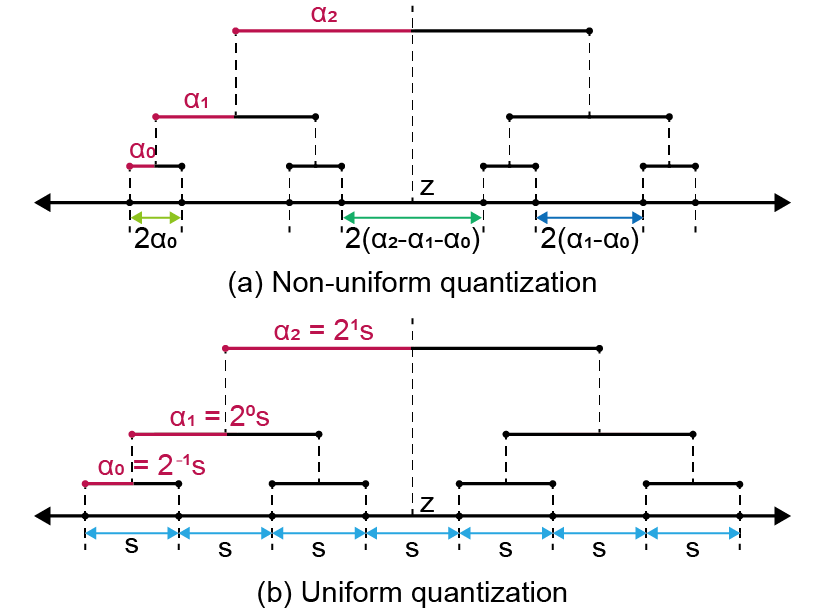}
  \caption{Extension of binary-coding quantization to support both non-uniform and uniform quantization formats, achieved by including a offset term ($q=3$).}
  \label{fig:bcq_extension}
\end{figure}

Binary Coding Quantization (BCQ) \cite{xu2018alternating} is a quantization technique that converts a real-valued weight value \( w \in \mathbb{R}\) into $q$-bit precision representation.
This is achieved by expressing $w$ as a linear combination of binary weights $\{b_i\}_{i=1}^q$ and their corresponding scaling factors $\{\alpha _i\}_{i=1}^q$, where $b_i \in \{ -1, 1 \}$ \cite{you2024shiftaddllm}.
The binary weights and scaling factors are optimized to minimize the quantization error using the following objective:
\begin{equation}
\underset{\alpha_i, b_i}{\arg \min} \left\| w - \sum_{i} \alpha_i b_i \right\|^2.
\end{equation}
Due to the absence of an analytical solution for minimizing the quantization error, heuristic methods are typically employed to approximate the optimal scaling factors and binary matrices.

While BCQ can be used for both activation and weight quantization, it leads to higher computational complexity when applied to activation quantization \cite{jeon2020biqgemm}.
Focusing on weight-only quantization to preserve accuracy, we will consider BCQ in the context of weight quantization. Consequently, the product of a FP activation vector $\textbf{x} \in \mathbb{R}^{n \times 1}$ and a weight matrix quantized with binary matrices $\textbf{B}_i \in \{ -1, 1 \}^{m \times n}$ and scaling factors $\alpha_i \in \mathbb{R}^{m \times 1}$ can be expressed as:
\begin{equation}
 \textbf{y} = \sum\limits_{i=1}^{q} (\alpha_i \circ (\textbf{B}_i \cdot \textbf{x}))
\end{equation}
where "\( \circ \)" denotes the Hadamard product.

Conventional BCQ, a non-uniform quantization method, is incompatible with most state-of-the-art models that use uniform quantization.
\cite{park2022lut} enhanced BCQ by incorporating an offset term $z$, thereby improving its representational capacity:
\begin{equation}
 w = \sum\limits_{i=1}^{q} ( \alpha _i \cdot b_i ) + z
\end{equation}
It has been demonstrated that uniform quantization can be effectively represented by appropriately adjusting the scaling factors and offset terms \cite{park2022lut}. 
As shown in Fig.~\ref{fig:bcq_extension}, conventional BCQ leverages multiple distinct scaling factors, while the addition of an offset term facilitates the representation of uniformly quantized values.

\begin{figure}[]
  \centering
  \includegraphics[width=0.9\linewidth]{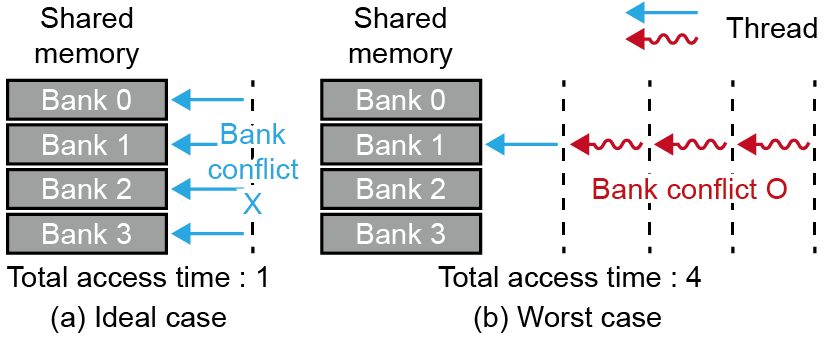}
  \caption{Comparison of bank conflicts during shared memory access.}
  \label{fig:bank_conflict}
\end{figure}

\begin{figure*}[]
  \centering
  \includegraphics[width=0.9\linewidth]{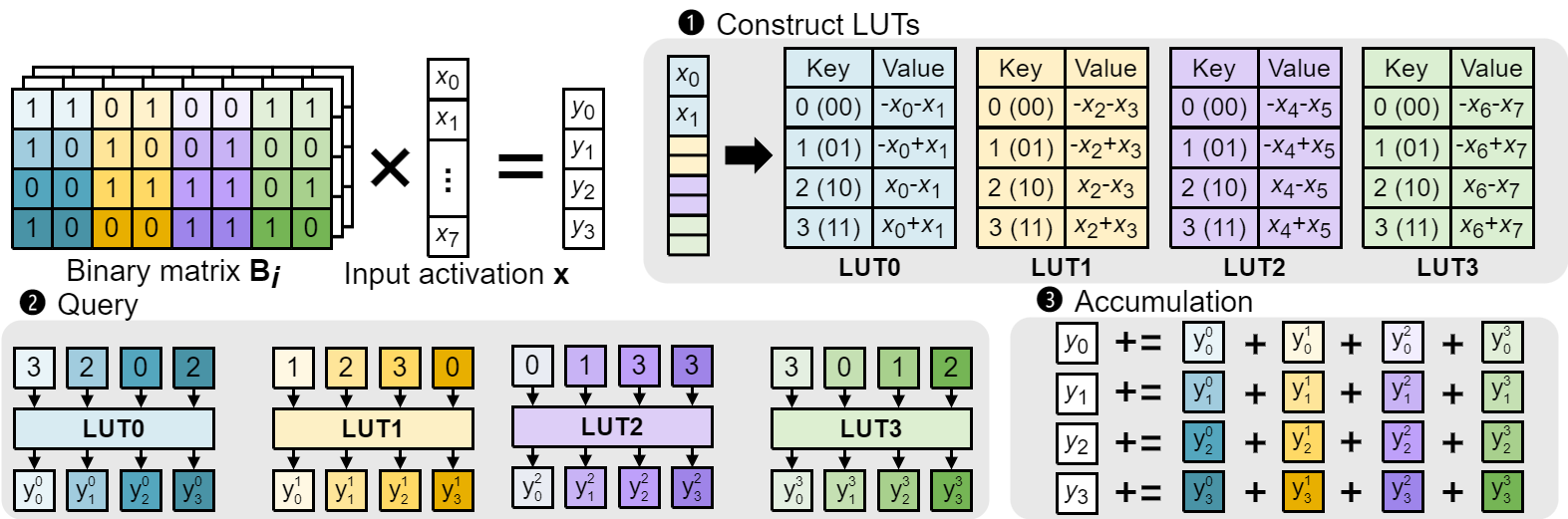}
  \caption{Illustration of look-up table based FP-INT GEMM.}
  \label{fig:overview}
\end{figure*}

\subsection{Bank Conflict}
To perform efficient parallel operations, it is necessary to access multiple data simultaneously. Many parallel processors achieve this by employing memory banking techniques, which split memory into banks, allowing multiple threads to access the memory concurrently. However, memory banking can encounter issues when the data required in a single operation cycle are located in the same bank, leading to bank conflicts \cite{gou2013addressing}.
This problem is particularly evident in GPUs, where a shared memory space is utilized to enable efficient parallel operations among threads. When different threads attempt to access the same bank simultaneously, bank conflicts arise. These conflicts hinder the achievement of predicted performance and lead to computational resource wastage, as operations that could be parallelized are instead executed serially.

In addition, bank conflicts pose a significant bottleneck when LUT-based methods are deployed on existing hardware platforms.
For instance, LUT-GEMM\cite{park2022lut} stores the LUTs in the shared memory of GPU and utilizes multiple threads for operations. 
During the LUT construction phase, bank conflicts are avoided as each thread is designed to access different banks in parallel. 
In contrast, during the LUT read phase, the randomness of the weight pattern often causes frequent bank conflicts, leading to performance degradation. 
Fig.~\ref{fig:bank_conflict} depicts the example of multiple threads accessing memory banks. 
In the worst-case scenario, all required accesses within a cycle are serialized, creating substantial performance overhead.
Therefore, for LUT-based operations to achieve optimal performance, it is essential to avoid bank conflicts.
This requires a novel LUT architecture that allows multiple parallel operators to simultaneously read different values without conflicts.

\section{Look-up Table-based FP-INT GEMM Accelerator}
In this section, we introduce FIGLUT, a novel computational methodology and accelerator design aimed at improving the efficiency of FP-INT GEMM operations by replacing computation with data retrieval from an LUT.
To address the issue of LUT bank conflicts that hinder parallel processing in existing LUT-based operations, we propose a specialized LUT structure.
This structure is designed to eliminate bank conflicts and enable efficient parallel access to LUT data.
We also analyze the impact of increased fan-out on power consumption in shared LUTs and propose an optimal processing element (PE) structure to maximize energy efficiency.
As a result, the FIGLUT design offers enhanced energy efficiency relative to conventional computing units.

\begin{table}[]
\centering
\caption{Example of Look-up Table when $\mu=3$ }
\label{tab:lut_example}
\begin{tabular}{ccc}
\hline
Binary Patterns & Key       & Value          \\ \hline
$\{-1,-1,-1\}$  & 0 (b'000) & $-x_1-x_2-x_3$ \\
$\{-1,-1,+1\}$  & 1 (b'001) & $-x_1-x_2+x_3$ \\
$\{-1,+1,-1\}$  & 2 (b'010) & $-x_1+x_2-x_3$ \\
$\{-1,+1,+1\}$  & 3 (b'011) & $-x_1+x_2+x_3$ \\
$\{+1,-1,-1\}$  & 4 (b'100) & $+x_1-x_2-x_3$ \\
$\{+1,-1,+1\}$  & 5 (b'101) & $+x_1-x_2+x_3$ \\
$\{+1,+1,-1\}$  & 6 (b'110) & $+x_1+x_2-x_3$ \\
$\{+1,+1,+1\}$  & 7 (b'111) & $+x_1+x_2+x_3$ \\ \hline
\end{tabular}
\end{table}

\subsection{Look-up Table based FP-INT GEMM}
Look-up tables are extensively employed to improve computational efficiency by substituting direct calculations with table look-ups, particularly when the results are confined to a predetermined set.
This technique can be extended to deep learning operations involving a binary weight matrix $\textbf{B} \in \{-1,+1\}^{M \times N}$, where output activations are calculated through simple addition and subtraction among the input activation vector $\textbf{x} \in \mathbb{R}^{N \times 1}$. 
Consequently, the output activations are confined to combinations of input activations, enabling the calculation and storage of all potential output values within the LUT. This approach substitutes actual computation with the retrieval of data from the LUT.

To construct an LUT, we introduce a hyperparameter $\mu$, which determines the number of binary weights used to form a key for retrieving a value from the LUT.
Note that as $\mu$ increases, the number of computations that can be replaced by a single LUT read also increases.
However, since the memory size required to store the LUT grows exponentially, finding the optimal value of $\mu$ is crucial.
Consider a binary matrix $\textbf{B} \in \{-1,+1\}^{4 \times 6}$ and an input activation vector $\textbf{x} \in \mathbb{R}^{6 \times 1}$:
\begin{equation}
\label{eq:matmul}
    \textbf{B} =
    \begin{bmatrix}
    +1 & -1 & -1 & -1 & -1 & +1 \\
    +1 & -1 & -1 & +1 & +1 & -1 \\
    -1 & +1 & -1 & -1 & -1 & +1 \\
    +1 & -1 & -1 & -1 & -1 & +1
    \end{bmatrix},
\end{equation}

\begin{equation}
    \textbf{x} =
    \begin{bmatrix}
    x_1 & x_2 & x_3 & x_4 & x_5 & x_6
    \end{bmatrix}^\top.
\end{equation}
For instance, in the computation of $\textbf{B}  \textbf{x}$ with $\mu=3$, we observe 
that operations such as $(x_1-x_2-x_3)$ and $(-x_4-x_5+x_6)$ are repeated multiple times.
These redundant computations increase as the size of the matrices grows with the increasing model size.
As shown in Table~\ref{tab:lut_example}, by pre-computing all possible combinations of inputs and storing them in the LUT, repeated computations can be avoided, allowing a single LUT read to replace multiple addition operations.
If $\mu$ is increased to 6, there are repeated computations such as $(x_1-x_2-x_3-x_4-x_5+x_6)$, which can also be efficiently managed using the LUT.
Although increasing $\mu$ enhances computational benefits, the memory space required to store the LUT increases exponentially.


Fig.~\ref{fig:overview} provides an overview of the LUT-based FP-INT GEMM process for computing $\textbf{B} \textbf{x}$ with $\mu=2$.
As previously mentioned, to reduce redundant computations and efficiently perform GEMM operations, LUTs are constructed using $\mu$ elements from the input activation \textbf{x}, customized for each input.
Each LUT contains $2^{\mu}$ values, with $\mu$ binary weights used as keys to retrieve each value.
During the query phase, $k$ RACs simultaneously access the shared LUT to retrieve their respective partial sum values.
In this example, $k=4$, indicates that four RACs can access a single LUT concurrently.
The specialized LUT structure for LUT-based GEMM, which enables concurrent access by $k$ RACs, will be detailed in the subsequent section.
Finally, the values retrieved from each LUT are stored in the accumulator.
Once this process is completed for all binary matrices $\textbf{B}_i$, the final output activation values are obtained.
Note that while the previous examples assume the input activation to be a vector for simplicity (corresponding to a GEMV operation), this approach is basically applied to GEMM operations by generating unique LUTs for each batch, allowing for efficient matrix computations.

\begin{figure}[]
  \centering
  \includegraphics[width=\linewidth]{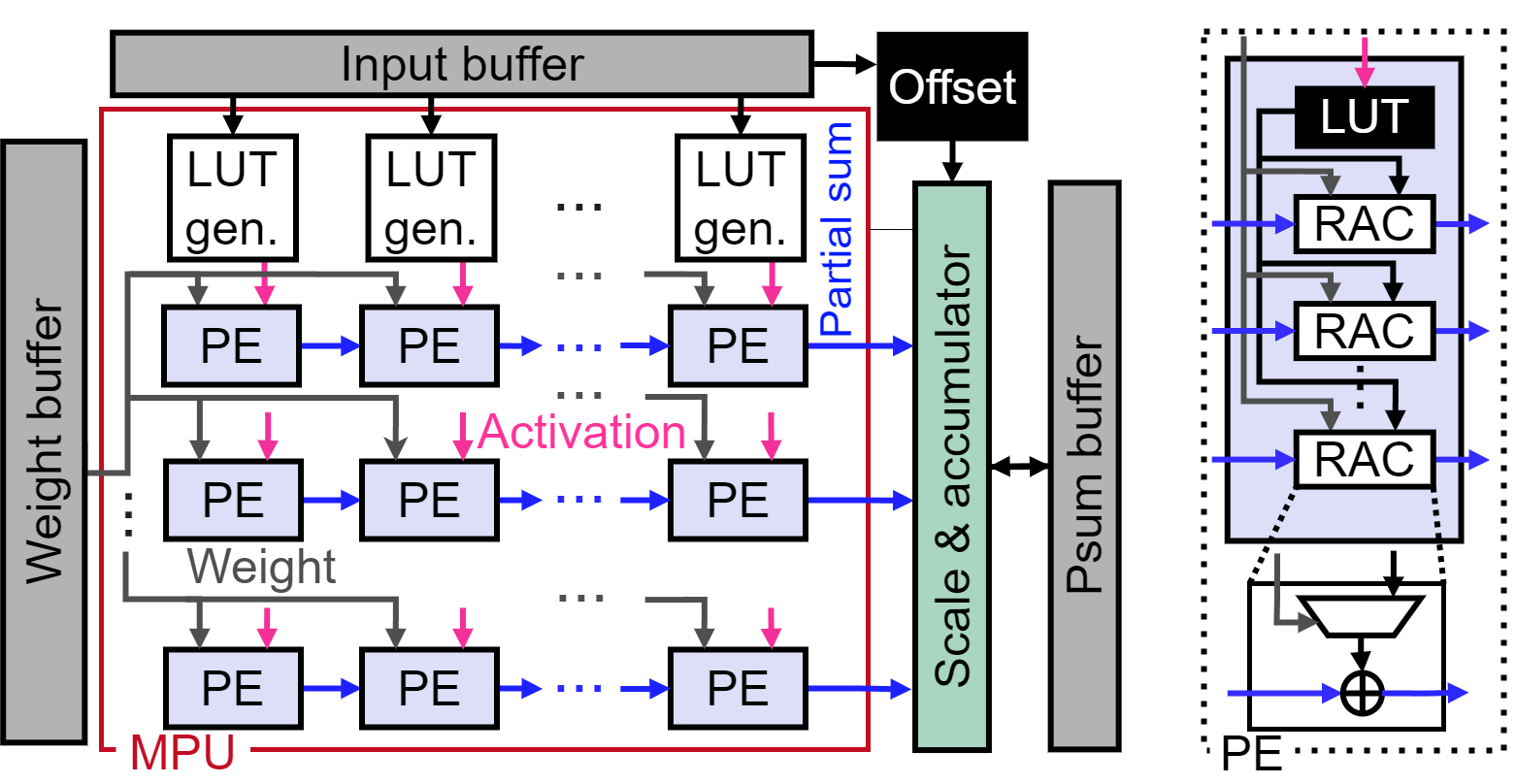}
  \caption{Overall MPU architecture of FIGLUT}
  \label{fig:overall_architecture}
\end{figure}

\subsection{Overall FIGLUT Architecture}

Fig.~\ref{fig:overall_architecture} depicts the overall architecture of the proposed LUT-based FP-INT GEMM accelerator, FIGLUT. 
In this section, we provide a detailed description of the matrix processing unit (MPU), which represents our primary contribution. 
A comprehensive overview of the entire system architecture will be discussed in Section~\ref{section:system_overview}.
FIGLUT leverages a 2D systolic array structure similar to that of the Google TPU \cite{jouppi2017datacenter}.
Inputs are fetched sequentially from the input memory through systolic data setup.
These input activations are fed into the LUT generators to construct the LUTs.
Each LUT generator takes $\mu$ input activations and generates all possible output partial sums on-the-fly.
These partial sums are then passed to the LUTs within the PEs, where they are used as LUT values.
To enable effective data reuse, LUT values from the PEs in each row are forwarded to the LUTs in the subsequent rows. 
Each PE houses $k$ RACs, and due to the unique structure of LUT, designed for LUT-specific operations, the $k$ RACs can read from the LUT concurrently. 
We perform an analysis to determine the optimal value of $k$, taking into account the fan-out overhead introduced when multiple RACs access the LUT at the same time. 
This analysis is elaborated in a later section.

FIGLUT employs a weight-stationary dataflow to optimize GEMM operations. Within each PE, the RAC includes a register that holds a $\mu$-bit weight pattern, serving as a key to access data from the LUT. The partial sums generated by each PE are incrementally gathered and subsequently sent to the PE in the next column. This iterative accumulation process takes place along each column of PEs until the ultimate partial sum is derived at the final PE. The eventual partial sum is multiplied by a scaling factor $\alpha_i$ and stored in the accumulator buffer. This cycle is performed for each bit plane, and once all bit planes are processed, the accumulated sums are summed with the offset value and routed to the output buffer.

\begin{figure}[]
  \centering
  \includegraphics[width=\linewidth]{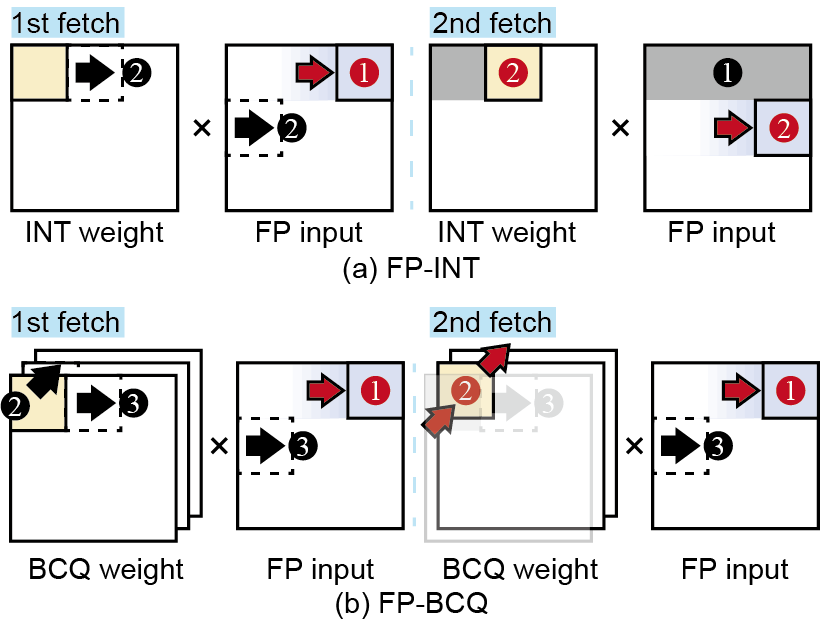}
  \caption{Illustration of the input and weight tile fetching sequence in (a) a systolic array accelerator for INT weight and (b) FIGLUT for BCQ weight. The arrows indicate the order in which the tiles are processed.}
  \label{fig:weight_fetch}
\end{figure}

Fig.~\ref{fig:weight_fetch} shows the fetching sequence of input and weight tiles.
For systolic array accelerators with INT weights, such as FIGNA~\cite{jang2024figna}, each weight has a single channel with multi-bit values. In contrast, FP-BCQ accelerators like FIGLUT and iFPU~\cite{kim2023winning} include $q$ binary weight bit planes.
In a weight-stationary dataflow, a weight tile is loaded once and reused, while inputs are loaded sequentially (\circled{1}) for both FP-INT and FP-BCQ accelerators.
However, the process diverges for FP-BCQ accelerators.
Instead of loading the next tile within the same bit plane, the weight tile for the next bit plane is loaded and processed (\circled{2} in Fig.~\ref{fig:weight_fetch} (b)). 
After processing all binary weight bit planes, the operation proceeds to the next tile, similar to the FP-INT accelerator.

\subsection{Conflict-free LUT Structure}

To efficiently perform LUT-based FP-INT GEMM operations, a specialized LUT structure is essential, as traditional memory structures are insufficient for the following requirements.
First, minimizing the power consumption associated with reading data from the LUT is crucial.
Since FIGLUT replaces computation with data retrieval, reducing the cost of data reads compared to the cost of computations enhances overall efficiency.
Second, the structure needs to support a large volume of simultaneous read and write operations.
Given the dynamic nature of activation values, the LUT should be reconfigurable on-the-fly, allowing concurrent writing of large amounts of data to update the LUT according to input activations. 
From a data retrieval perspective, multiple RACs operate in parallel, each using different weight patterns as keys to access data in the LUT. 
Therefore, the architecture should facilitate parallel data access while avoiding bank conflicts.

\begin{figure}[t]
  \centering
  \includegraphics[width=1\linewidth]{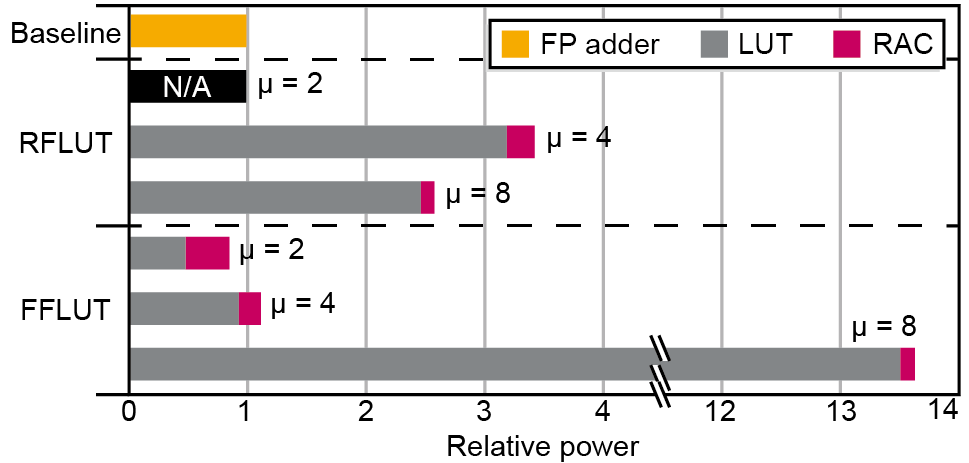}
  \caption{Relative Power Consumption of RFLUT and FFLUT Compared to FP Adder Baseline Across Different $\mu$ Values}
  \label{fig:RFLUT_FFLUT_power}
\end{figure}

To ensure that retrieving precomputed results from an LUT consumes less power than using FP adders, we measure the power consumption of an FP adder as a baseline. 
We then compare this with the power consumption of LUT-based operations.
Fig.~\ref{fig:RFLUT_FFLUT_power} demonstrates the power comparison between FP adders and various LUT implementations across different $\mu$ values, while maintaining equivalent throughput.
To address the targeted hardware constraints, we implemented two types of LUTs: (1) a traditional Register File based LUT (RFLUT) and (2) a proposed Flip-Flop based LUT (FFLUT).

The RFLUT, constructed using register files commonly employed in CPU register arrays, uses a straightforward approach where the weight pattern serves as the read register number.
For the comparison, RFLUT blocks are generated using a 28nm memory compiler, allowing us to customize the design to target low-power LUT-based GEMM operations.
The generated register file is then integrated into the hardware design as an IP macro block.
All power analyses of LUT structures, including FFLUT, are tested based on the physical layout obtained after the P\&R process in ICC2.
As the RFLUT with $\mu=2$ is too small to be generated by the compiler in this environment, it is not considered for measurement.
Generally, RFLUTs exhibit higher read power consumption compared to FP addition operations.
While an RFLUT with $\mu=4$ requires less power per read than one with $\mu=8$, it needs twice the number of reads, leading to higher overall power consumption.
As a result, RFLUT is not suitable for the LUT-based method, and a novel method is necessary to construct LUTs.

\begin{figure}[]
  \centering
  \includegraphics[width=1.0\linewidth]{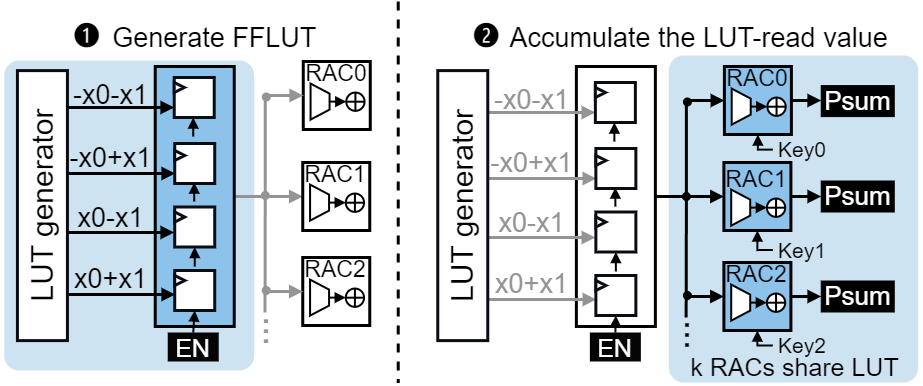}
  \caption{Architecture of the Flip-Flop based Look-Up Table (FFLUT)}
  \label{fig:FFLUT}
\end{figure}

To overcome the limitations of traditional RFLUTs, we propose a Flip-Flop-based Look-Up Table (FFLUT) structure, illustrated in Fig.~\ref{fig:FFLUT}.
The FFLUT consists of a bundle of flip-flops accessed through multiplexers.
Each enabled flip-flop outputs consistent values once updated during the LUT generation phase, maintaining this output until it is reset.
This design allows the processing unit to retrieve values by multiplexing the corresponding weight keys to access the required flip-flops.
Unlike RFLUTs, which are constrained by their limited read/write ports, the FFLUT enables simultaneous access by multiple units without encountering bank conflicts. 
This is achieved by equipping the FFLUT with dedicated multiplexers that allow different units to access distinct keys concurrently. 
As a result, multiple units can share a single LUT efficiently, significantly reducing the overhead associated with LUT-based operations and improving energy efficiency.

To further optimize the data retrieval process in FFLUTs, we introduce a specialized operator called the RAC unit, designed as a replacement for the conventional MAC unit.
Each PE in FIGLUT is equipped with a single LUT and multiple RAC units, as shown in Fig.~\ref{fig:overall_architecture}. 
Unlike the MAC unit, which performs a multiplication followed by an accumulation into a partial sum, the RAC unit retrieves values directly from the LUT using a weight pattern as a key and accumulates them into the partial sum. 
Hence, the RAC unit can execute LUT-based operations more effectively, enhancing the overall efficiency of the FIGLUT architecture.

\begin{figure}[]
  \centering
  \includegraphics[width=1\linewidth]{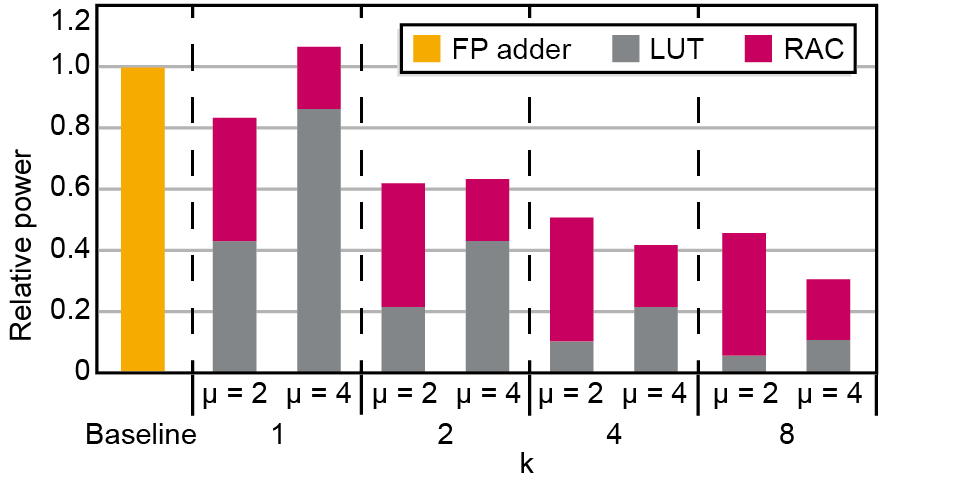}
  \caption{Relative power comparison of baseline, $\mu=2$ and $\mu=4$ for various the number of RACs per LUT configuration.}
  \label{fig:rac_num}
\end{figure}

To achieve energy-efficient GEMM operations, our approach incorporates an optimal architecture search that allows multiple RAC units to share a single LUT. 
This optimization process involves two key steps. 
First, we determine the optimal $\mu$ value for the LUT, balancing the gains of the computation against the power consumption of the LUT. 
Second, based on the chosen $\mu$ value, we identify the optimal number of RACs per LUT by evaluating the impact of fan-out on power consumption.
We introduce a variable $k$ to represent the LUT fan-out within a single PE, denoting the number of RACs sharing a single LUT.
Fig.~\ref{fig:rac_num} shows the power consumption of FIGLUT with $\mu$ values of 2 and 4, varying $k$. 
The baseline for relative power measurements is set using the results of the same operations performed with FP adders.
Due to the significantly large size and high power consumption of LUTs for $\mu=8$, as shown in Fig.~\ref{fig:RFLUT_FFLUT_power}, this configuration is excluded from consideration.
As $k$ increases, the total number of LUTs decreases by sharing an LUT with multiple RACs, leading to a gradual reduction in LUT power consumption. 
Note that the total number of RACs required for computation remains constant regardless of $k$, as it depends solely on the $\mu$ value.
This is because the overall number of read-accumulate operations is unaffected by the sharing configuration.
However, smaller $\mu$ values require more RACs, resulting in higher RAC power consumption for $\mu=2$ compared to $\mu=4$.
When LUTs are not shared and are exclusively used by a single RAC ($k=1$), the larger LUT size for $\mu=4$ leads to higher power consumption, resulting in higher total power consumption compared to $\mu=2$.
However, as $k$ increases and LUTs are shared among multiple RACs, the number of LUTs decreases, reducing overall power consumption.
Based on these findings and considering sufficiently large $k$ values in our architecture, we use the FIGLUT architecture with $\mu=4$ in this paper.

\begin{figure}[]
  \centering
  \includegraphics[width=1\linewidth]{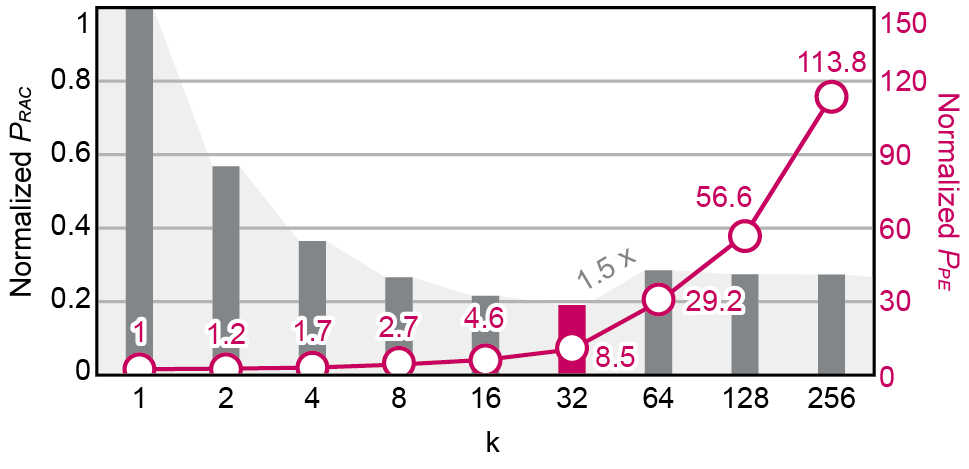}
  \caption{$P_{RAC}$ and $P_{PE}$ analysis for various RAC numbers. Normalized by $k=1$ results.}
  \label{fig:fanout}
\end{figure}


Managing fan-out is a critical challenge in designing an energy-efficient accelerator capable of sharing a single LUT among multiple RACs. In digital circuits, fan-out occurs when a signal is distributed to multiple circuit elements, leading to increased power use and signal delay. In the context of FIGLUT, as multiple RACs access the LUT simultaneously, the fan-out effect requires additional power to deliver signals to each RAC efficiently.
While sharing a single LUT among multiple RACs reduces the total number of LUTs, it also increases the power consumption of each LUT due to the heightened fan-out. This trade-off necessitates careful management to achieve an optimal balance between the benefits of reducing the total number of LUTs and the drawbacks of increased power demand from fan-out. Effectively addressing this issue is crucial for minimizing overall power use while maintaining energy efficiency in LUT-based GEMM operations.

Fig.~\ref{fig:fanout} shows the power consumption of the PE ($P_{PE}$) and each RAC ($P_{RAC}$) as a function of LUT fan-out.
To measure the impact of fan-out, all experimental results are obtained using the physical layout of the PEs (including LUTs and RACs) after the P\&R process.
The power consumption per RAC unit $P_{RAC}$ is calculated by dividing $P_{PE}$ by $k$.
As $k$ increases, the number of RACs within the PE increases, resulting in higher overall PE power.
Initially, increasing $k$ reduces $P_{RAC}$ because the power consumption of the shared LUT is distributed across the $k$ RACs.
However, as $k$ continues to grow, the fan-out overhead becomes significant, increasing $P_{RAC}$.
Based on the observation, we determine the optimal value of $k$ to be 32 for configuring PEs in our architecture.
Thus, each PE in our architecture includes one shared LUT and 32 RACs.

\subsection{Optimizing LUT using Vertical Symmetry}

\begin{table}[]
\centering
\caption{Comparison of relative power consumption of LUT and other components}
\label{tab:hfflut}
\begin{tabular}{ccccc}
\hline
       & LUT   & MUX   & Decoder & MUX+Decoder \\ \hline
FFLUT  & 1.000 & 0.003 & 0.000   & 0.003       \\
hFFLUT & 0.494 & 0.002 & 0.003   & 0.005       \\ \hline
\end{tabular}
\end{table}

In LUT-based FP-INT GEMM, the primary overhead compared to conventional GEMM operations is the LUT itself. 
Reducing LUT power consumption is therefore critical for enhancing the efficiency of FIGLUT. 
The FFLUT is composed of flip-flops, with power consumption scaling linearly with its size. 
To address this, we propose a half-FFLUT (hFFLUT) technique, which reduces the LUT size by half.

\begin{figure}[]
  \centering
  \includegraphics[width=\linewidth]{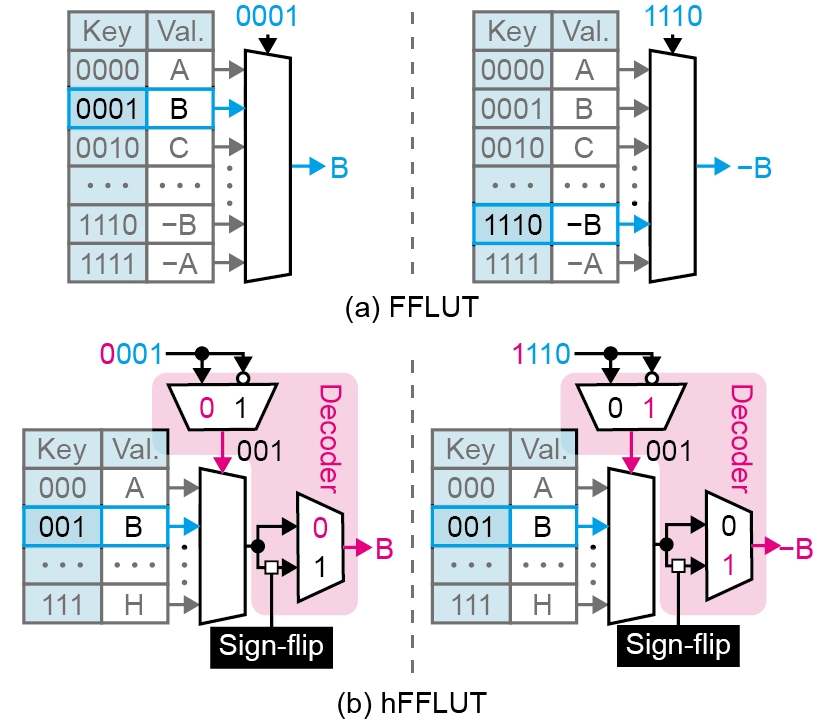}
  \caption{Block diagram of the (a) FFLUT and (b) hFFLUT with proposed decoder.}
  \label{fig:hFFLUT}
\end{figure}

As shown in Table~\ref{tab:lut_example}, LUT values exhibit vertical symmetry.
In other words, since the LUT is composed of the calculation results of $\mu$ FP values through combinations of addition and subtraction, for every combination, there exists a corresponding value that can be directly obtained by simply flipping the sign.
By exploiting this characteristic, we store only the upper half of the table and use the key to decode the final LUT value.
Fig.~\ref{fig:hFFLUT} presents a block diagram illustrating the hFFLUT and the proposed decoding technique when $\mu=3$.
The most significant bit (MSB) of the key is used as a select signal to choose the $(\mu-1)$-bit key for the hFFLUT.
Then, the final value read from the hFFLUT is then sign-flipped based on the MSB to obtain the complete value.

Although the hFFLUT effectively reduces the size of the LUT, it introduces additional decoding overhead.
Table~\ref{tab:hfflut} evaluates the hardware complexity of the proposed decoding circuit for hFFLUT compared to the FFLUT multiplexer used to fetch data from the LUT.
The decoding circuit for hFFLUT, including the multiplexer, demonstrates higher power consumption than the multiplexer used in the FFLUT.
However, since LUT power is dominant in the overall power consumption, the additional overhead of the decoding circuit is relatively trivial in comparison.
Consequently, the proposed hFFLUT effectively halves the power consumed by the LUT in the entire system, thereby significantly enhancing the energy efficiency of FIGLUT.

\subsection{Efficient LUT Generator}

\begin{figure}[t]
  \centering
  \includegraphics[width=1\linewidth]{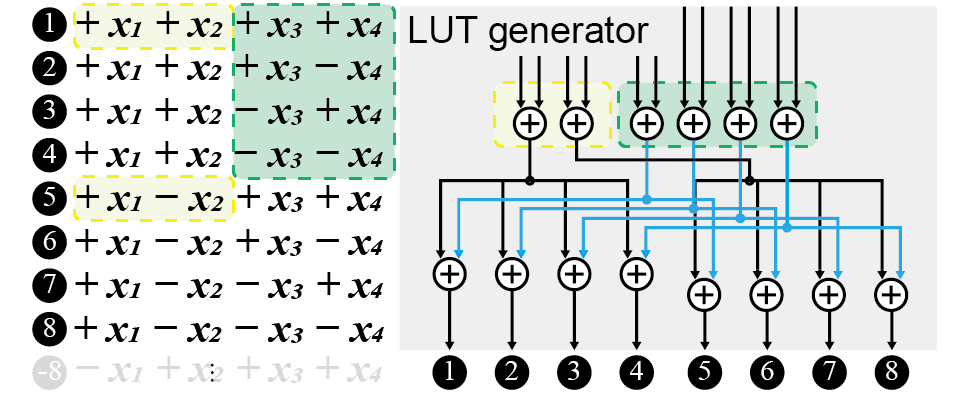}
  \caption{Required generating pattern and LUT generator module for $\mu=4$.}
  \label{fig:LUTgen}
\end{figure}

An efficient table generator that dynamically adapts to changing input activations is essential to calculate LUT elements.
We implemented a two-step tree-like hardware module that changes the sign of inputs and sums them up in parallel, generating multiple LUT values simultaneously. 
By leveraging the hFFLUT, which halves the number of LUT elements, the generator is designed to pre-compute results for only half the patterns.
As shown in Fig.~\ref{fig:LUTgen}, the lower bit patterns repeat for specific upper bit patterns (yellow shaded), allowing us to minimize the addition operations for the upper bit patterns. 
Duplicated lower bit parts (green shaded) are calculated once and connected in parallel to the next adders. 
By combining the results of upper 2-bit patterns with the lower 2-bit patterns, we can generate all the necessary hFFLUT patterns.
This approach reduces the number of adders and the total addition operations by 42\% compared to a straightforward implementation of a LUT generator. 
Assuming all LUT elements are used, for $\mu=4$, the LUT generator requires 14 additions to compute the complete set of results. 
In contrast, straightforward hardware without an LUT requires $\mu - 1 = 3$ additions per result. 
Thus, for $k>4$, the proposed LUT generator performs fewer additions to compute the same number of results compared to straightforward hardware with $k$ RACs. 
As $k$ increases, the addition saving of the proposed method becomes even more prominent, enabling a more efficient LUT generation.

\subsection{System overview}
\label{section:system_overview}

\begin{figure}[]
  \centering
  \includegraphics[width=1\linewidth]{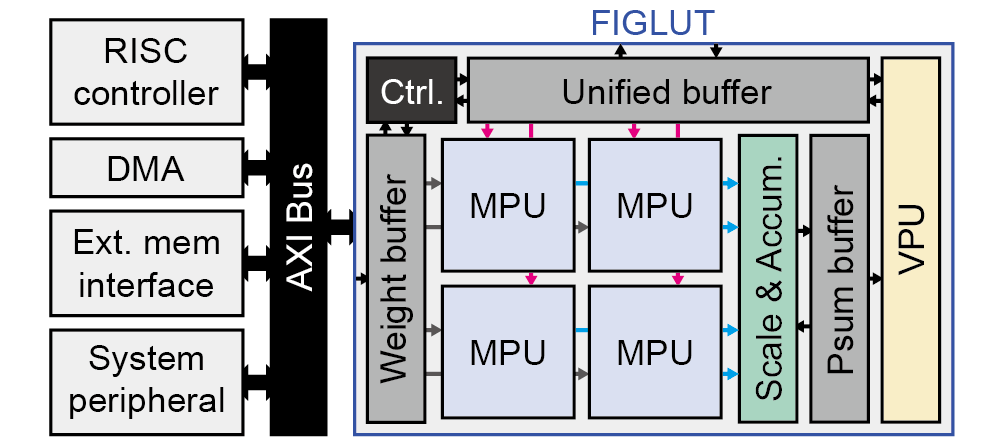}
  \caption{System architecture of FIGLUT.}
  \label{fig:system_overview}
\end{figure}

Fig.~\ref{fig:system_overview} provides an overview of the entire system, including FIGLUT. 
As a hardware accelerator module, FIGLUT is designed to be flexible and is not limited to specific system configurations, enabling seamless integration into a wide range of system architectures beyond the proposed setup.
The system adopts a shared memory structure between the host and FIGLUT, rather than using separate memory spaces.
Data transfer between the host and FIGLUT is facilitated through the AXI bus. 
This architecture eliminates the need to explicitly transfer final computation results from FIGLUT to the host; instead, the host can directly access the shared memory to retrieve the results efficiently.
It minimizes the overhead associated with data movement between FIGLUT and the host, thereby maximizing overall system efficiency.

FIGLUT comprises on-chip buffers for storing computation data, along with MPUs and a vector processing unit (VPU) to efficiently manage non-GEMM operations.
To maximize efficiency, FIGLUT adopts a tile-based GEMM operation approach, where the data required for each tile computation is stored in the on-chip buffer using double buffering techniques.
The MPU executes LUT-based GEMM operations, storing the intermediate computation results as partial sums in the Psum buffer for subsequent usage.
The final GEMM results are then processed by the VPU, and the output data is stored in a unified buffer.
This approach minimizes off-chip memory access and maximizes system efficiency.

\begin{figure*}[]
  \centering
  \includegraphics[width=\linewidth]{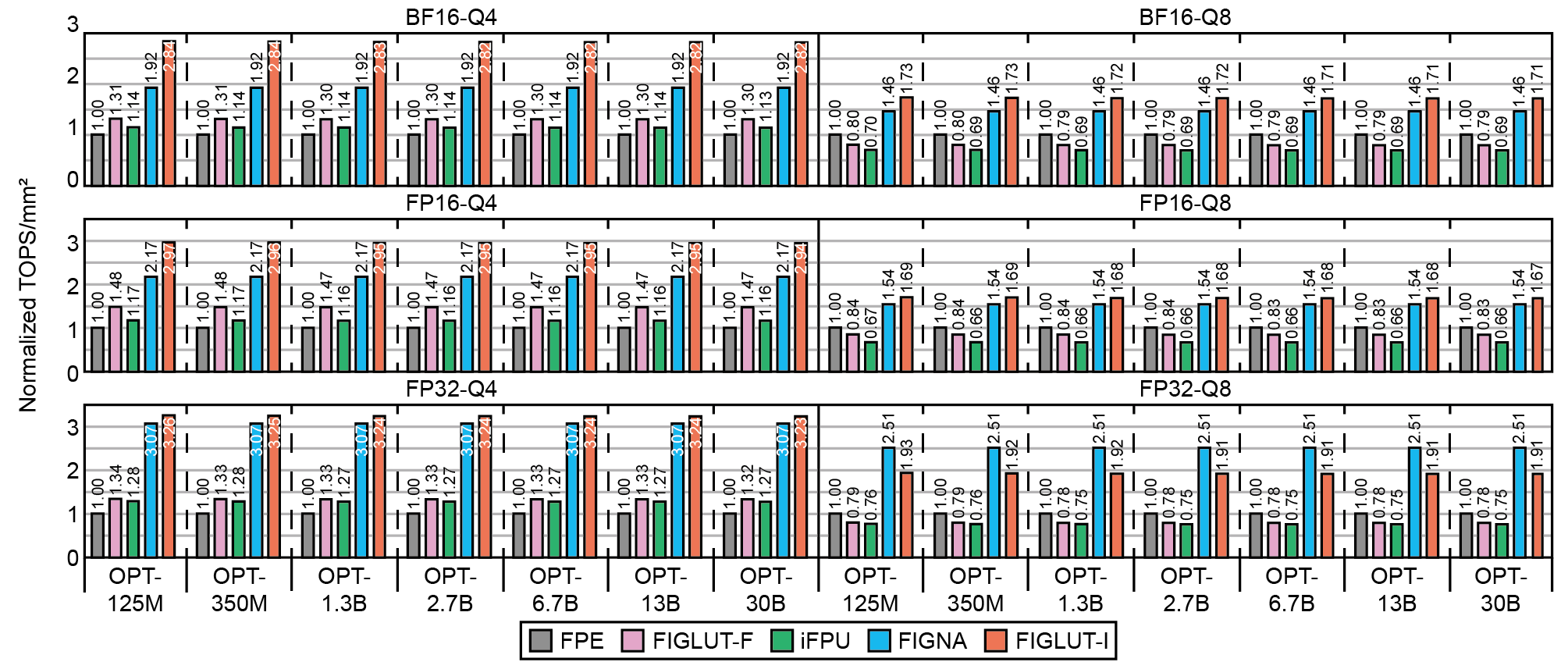}
  \caption{TOPS/$mm^2$ of hardware engines for Q4 and Q8 language models, normalized by FPE results.}
  \label{fig:tops_mm2}
\end{figure*}

\section{Evaluation}

\subsection{Accuracy Evaluation}

\begin{table}[]
\centering
\caption{Perplexity results of the OPT family models using different GEMM engines.
}
\label{tab:wiki_perplexity}
\begin{tabular}{lcccccc}
\hline
OPT     & \multicolumn{1}{c}{350M} & \multicolumn{1}{c}{1.3B} & \multicolumn{1}{c}{2.7B} & \multicolumn{1}{c}{6.7B} & \multicolumn{1}{c}{13B} & \multicolumn{1}{c}{30B} \\ \hline
GPU        & 55.24                    & 67.95                    & 35.46                    & 24.13                    & 20.93                   & 19.17                   \\
FIGLUT-F   & 55.24                    & 67.95                    & 35.46                    & 24.13                    & 20.93                   & 19.17                   \\
FIGLUT-I   & 55.24                    & 67.95                    & 35.46                    & 24.13                    & 20.89                   & 19.17                   \\ \hline
\end{tabular}
\end{table}

To evaluate the numerical accuracy of the proposed FIGLUT, we conduct an inference accuracy assessment using both an NVIDIA GPU and the FIGLUT engine. 
To further optimize our engine, we utilize the pre-alignment technique \cite{kim2023winning, jang2024figna} used in FP-INT GEMM to FIGLUT, resulting in two variations: FIGLUT-F, which bypasses the pre-alignment technique, and FIGLUT-I, which incorporates it.
We evaluate the transformer-based OPT model family \cite{zhang2022opt} on the WikiText-2 dataset \cite{merity2016pointer} for a language modeling task, measuring perplexity as the metric.
All models use FP16 activations and 4-bit weights, with the weights quantized using the simple uniform quantization method, round-to-nearest (RTN).
Note that we use FP32 for accumulation to preserve the accuracy of the accumulated results \cite{burgess2019bfloat16, henry2019leveraging}.
Table~\ref{tab:wiki_perplexity} shows the perplexity results to compare the numerical accuracy of each GEMM engine against the NVIDIA GPU.
Although the order of FP operations may change, potentially leading to differences in values, the results for FIGLUT-F show no significant accuracy loss compared to the NVIDIA GPU, thanks to the higher precision (FP32) accumulation.
FIGLUT-I also demonstrated comparable numerical accuracy with pre-alignment method, consistent with findings reported in previous studies \cite{kim2023winning, jang2024figna}.

\begin{figure}[]
  \centering
  \includegraphics[width=\linewidth]{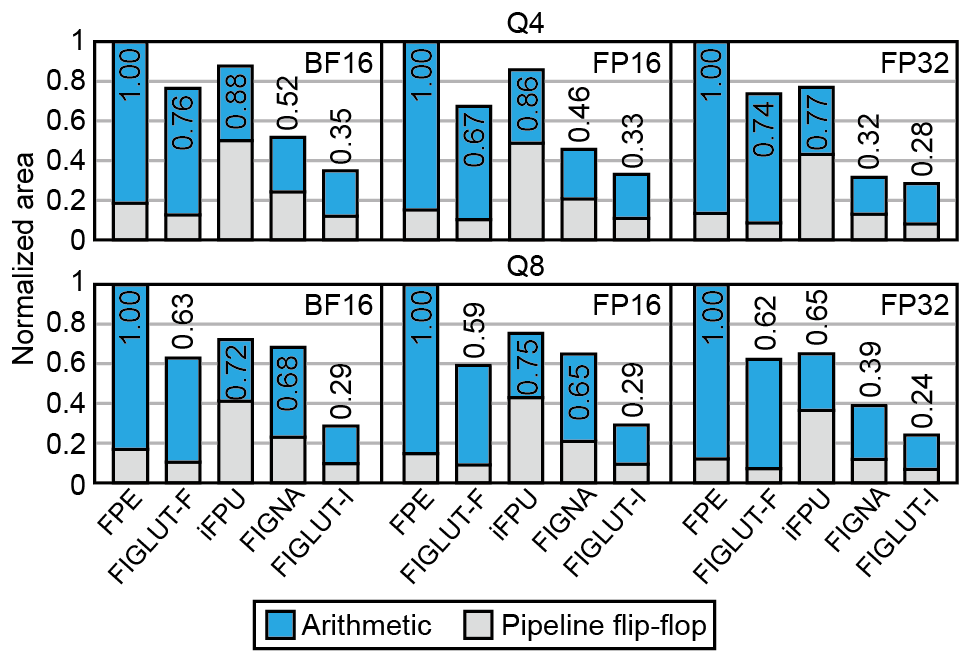}
  \caption{Area breakdown of MPUs for six variants of input formats normalized by results of FPE.}
  \label{fig:area_breakdown}
\end{figure}

\begin{figure*}[]
  \centering
  \includegraphics[width=1.0\linewidth]{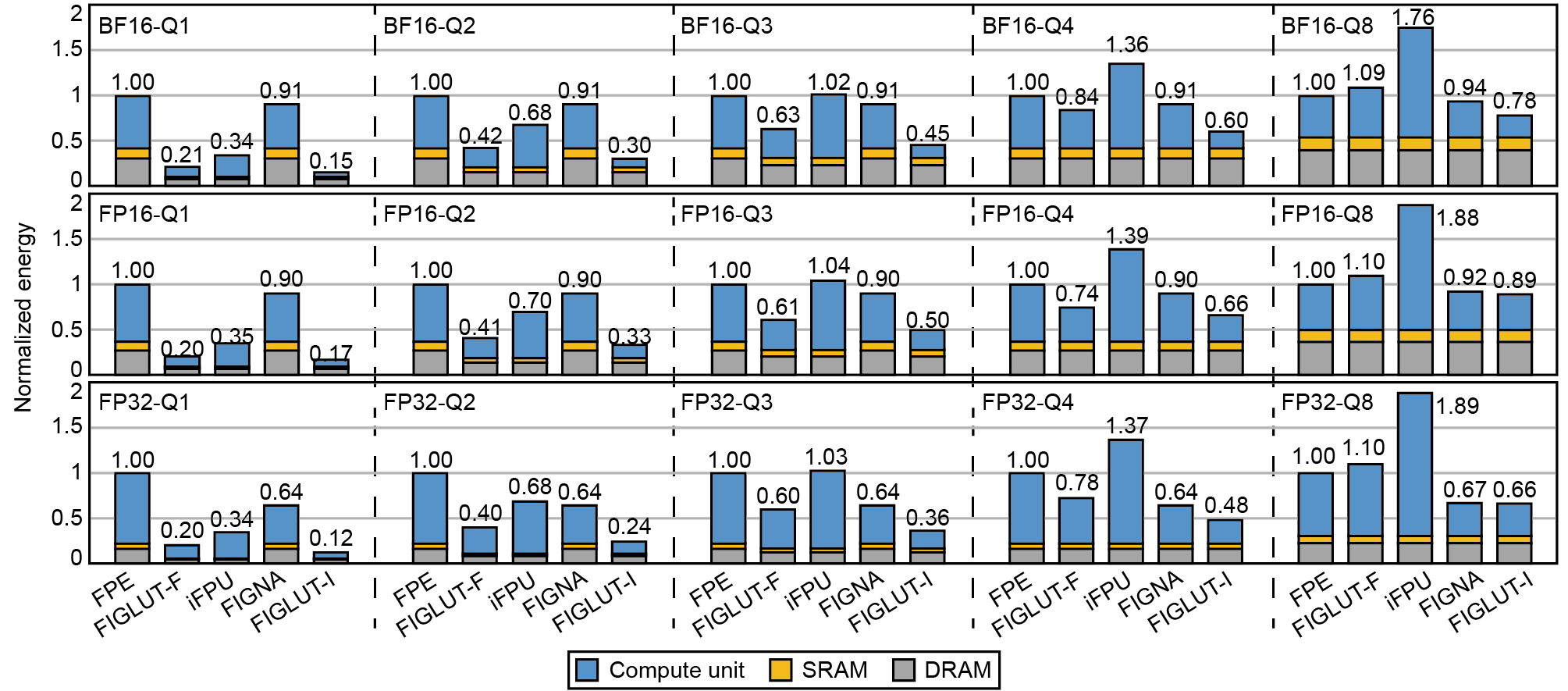}
  \caption{Normalized energy breakdown of FP-INT GEMM hardware engines across different bit precisions.}

  \label{fig:energy_breakdown}
\end{figure*}

\subsection{Hardware Evaluation}
\paragraph{Configuration Setup}
We evaluate five types of hardware engines: FIGLUT-F, FIGLUT-I, FPE, iFPU \cite{kim2023winning}, and FIGNA \cite{jang2024figna}.
The FPE serves as the baseline hardware engine, featuring PEs that dequantize INT weights to FP format with the same precision as the activation inputs, followed by FP multiplication and FP accumulation.
For both iFPU and FIGNA, the engines carry out pre-alignment procedures to transform FP input activations into integer mantissas with a common exponent prior to performing MAC operations.
For comparative analysis with the engines, the proposed FIGLUT hardware is implemented in two versions with different processing data formats: FIGLUT-F, which uses FP inputs directly into the LUT element, followed by FP RAC, and FIGLUT-I, which performs integer operations after pre-alignment. 
All hardware engines are designed to handle inputs with the same bit precision and word size and produce outputs in the same manner. 
Notably, iFPU and FIGLUT support the BCQ format, whereas FPE and FIGNA utilize the INT format.
To conveniently reference the different weight formats each hardware uses, we denote them by their bit-widths, using a notation such as Q4 for both 4-bit BCQ and INT weight.

For fair comparison, all engines are designed with MPUs and VPUs that provide identical throughput. 
Specifically, for Q4 weight inputs, FPE and FIGNA have $64 \times 64$ PE arrays, while iFPU processing 1-bit weight employs a $64 \times 64 \times 4$ array. 
FIGLUT uses an $2 \times 16 \times 4$ array, which, considering $\mu=4$ and $k=32$, results in the same number of computational units as iFPU.
For weights with a bit precision smaller than Q4, the same hardware configurations are used. 
For Q8 weights, results are obtained using extended versions of the FPE and FIGNA hardware, where the input weight bit precision was expanded to 8-bit. 
To leverage the bit-serial characteristics of iFPU and FIGLUT, both hardware engines are configured to match the throughput of Q4, affecting the operation cycles required for matrix multiplication for other weight precision.

The hardware engines are synthesized with Synopsys Design Compiler for 100MHz operation with 28nm CMOS technology.
Basic FP and integer modules, including the integer-to-FP converter in FPE, use Synopsys DesignWare components. 
All input, weight, and output buffers are composed with SRAM in 28nm CMOS technology, with off-chip DRAM results derived from the CACTI simulator \cite{vais2017cacti}. 
Data from the external DRAM is transferred directly to the FIGLUT accelerator SRAM buffer under the management of the host controller.
The data is subsequently utilized by the MPU and VPU for computations.
Energy metrics, including power consumption and latency associated with memory data transfers, are incorporated into all evaluation results, with the exception of area measurements, as the DRAM area is unspecified.

\paragraph{Area Evaluation}

Fig.~\ref{fig:area_breakdown} illustrates the normalized area breakdown of the MPUs across varying engine types, considering different input precision scenarios.
The breakdown is categorized into two primary components: arithmetic logic, responsible for executing computations, and flip-flops to temporarily store values.
The result shows that the arithmetic part occupies a larger area in FPE and FIGLUT-F compared to other integer-operating engines, with FIGLUT-F having a smaller area than FPE because it performs FP addition instead of FP multiplication.
In the Q8 results, the arithmetic area of FIGNA increases more significantly compared to FPE because the computational units in FIGNA scale with the bit precision of the weights, whereas FPE only encounters an increase in the overhead of the dequantization module without a change in the actual input bit-width of FP operators.
When comparing FIGLUT-I and FIGNA, despite the addition of an LUT generator, the use of LUT-based operations reduces the number of computational units, resulting in a similar arithmetic area as FIGNA.
Additionally, the introduction of LUT-based operations reduces the overall flip-flop area compared to other hardware architectures.
With a smaller $2 \times 16 \times 4$ MPU, compared to the conventional configurations of $64 \times 64$ and $64 \times 64 \times 4$, FIGLUT effectively reduces the number of pipelining stages required for systolic array operations.
Furthermore, while traditional hardware designs may demand up to 63-stage input buffers for weight-stationary operations, FIGLUT requires a maximum of only 15-stage input buffers, enhancing both efficiency and resource utilization.

\begin{figure*}[]
  \centering
  \includegraphics[width=\linewidth]{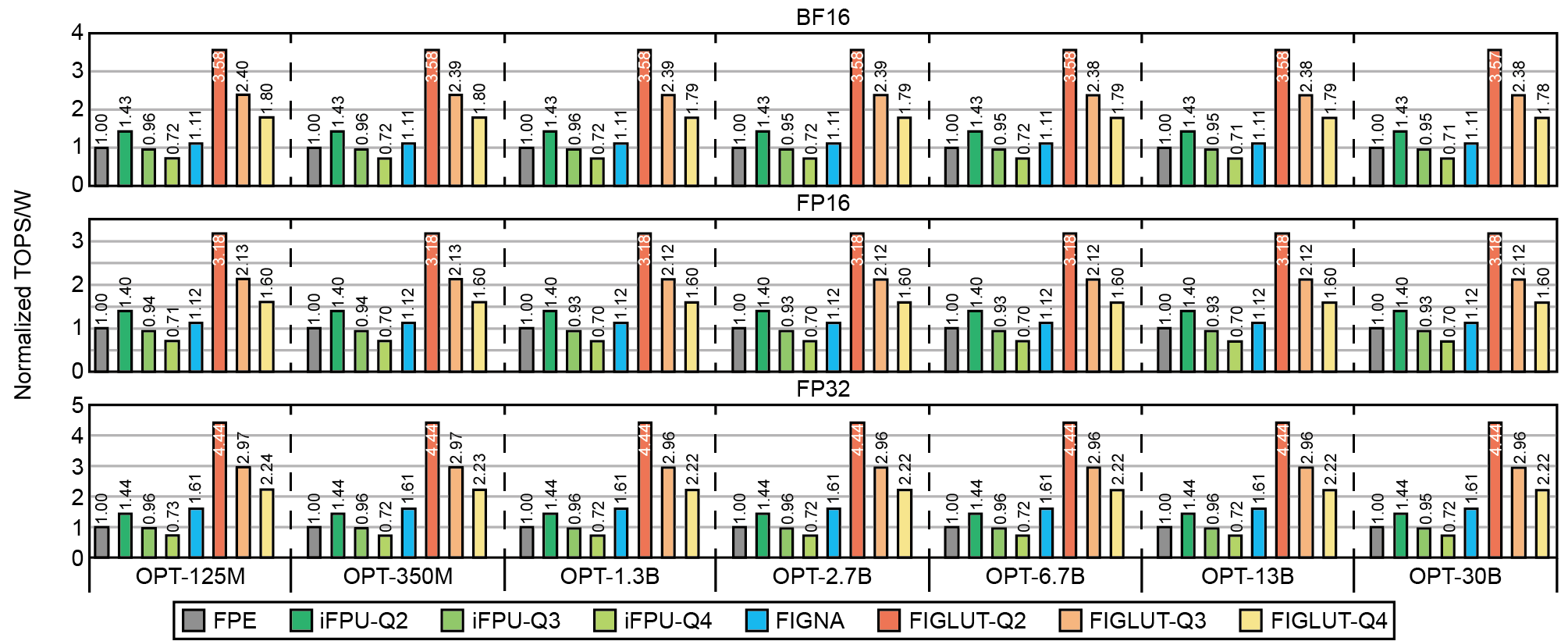}
  \caption{TOPS/W of hardware engines for under 4-bit OPT language models, normalized by FPE results.}
  \label{fig:tops_w}
\end{figure*}

Fig.~\ref{fig:tops_mm2} presents the area efficiency targeting various LLMs ranging from OPT-125M to OPT-30B. 
Excluding hardware designs with bit-serial architecture, the overall trends inversely follow the area breakdown shown in Fig.~\ref{fig:area_breakdown} across all network and input bit-widths. 
For instance, as previously noted, the PE size of the FIGNA is relatively larger in Q8 compared to Q4, resulting in reduced normalized energy efficiency. 
However, the bit-serial engines exhibit different results because TOPS/$mm^2$ reflects the time required for computations.
With the same hardware configuration, hardware designs with bit-serial architecture consume approximately twice the cycles with increased weight bit-width, leading to more significant performance degradation in Q8. 
The difference between FIGNA and FIGLUT-I decreases for FP32 due to the increased mantissa bit-width, contributing to performance reversal in FP32-Q8.
Nonetheless, the proposed engines achieve up to $1.5 \times$ higher area efficiency than state-of-the-art performance in the current trend of sub-4-bit weight-only quantization.
Note that non-GEMM operations cause slight variations by model size, but their impact is minimal as GEMM operations dominate the workload.

\paragraph{Energy Evaluation}

Fig.~\ref{fig:energy_breakdown} shows the energy breakdown for each input data type pair, normalized to the energy consumption of the FPE.
To assess and compare the energy efficiency across various bit precisions—including Q4, Q8, and sub-Q4 precisions (Q1, Q2, Q3)—we conduct simulations using the OPT-6.7B model.
As previously mentioned, all systems are configured to have the same maximum performance to ensure a fair comparison.
In Q8 operations, while bit-serial hardware continues to utilize identical architecture irrespective of bit precision, fixed precision architectures like FPE and FIGNA are modified to support 8-bit weights.
For sub-4-bit weight precision, the number of bit-serial operations decreases proportionally as the bit-width is reduced, whereas fixed bit-width hardware needs to process sub-4-bit data in padded 4-bit format for computation.
This allows bit-serial hardware such as iFPU and FIGLUT to complete computations more quickly, effectively reducing energy consumption as weight precision decreases.
The energy consumption of the compute unit, including the MPU and the VPU, for Q4 precision follows a trend similar to the area breakdown depicted in Fig.~\ref{fig:area_breakdown}.
However, iFPUs, which employ a greater number of flip-flops than FPEs, suffer from higher power consumption despite being more area-efficient, resulting in a distinct energy profile.
In the case of FIGLUT-I, the size of the PE increases from BF16 to FP32 because the computational units scale proportionally with the mantissa of the input activation.

Fig.~\ref{fig:tops_w} illustrates the power efficiency of the hardware engines when performing sub-4-bit weight operations on LLMs of various sizes. 
To highlight the performance variations in bit-serial operation engines, we present power efficiency metrics for Q2, Q3, and Q4 operations. 
Given that FIGLUT-I shows better power efficiency with integer operations compared to FIGLUT-F, we focus solely on FIGLUT-I for this evaluation.
As expected, reducing the weight bit-width decreases the number of cycles required for computations in bit-serial hardware, thereby improving energy efficiency.
FIGLUT consistently achieves the highest TOPS/W across all weight bit-widths, with FIGLUT-Q2 demonstrating particularly superior performance gain compared to other configurations.
This result explains that in memory-bound computing environments, i.e., LLMs, minimizing the weight bit-width and leveraging bit-serial operations can significantly enhance computational speed, resulting in substantially higher power efficiency compared to traditional hardware.

\begin{figure}[]
  \centering
  \includegraphics[width=\linewidth]{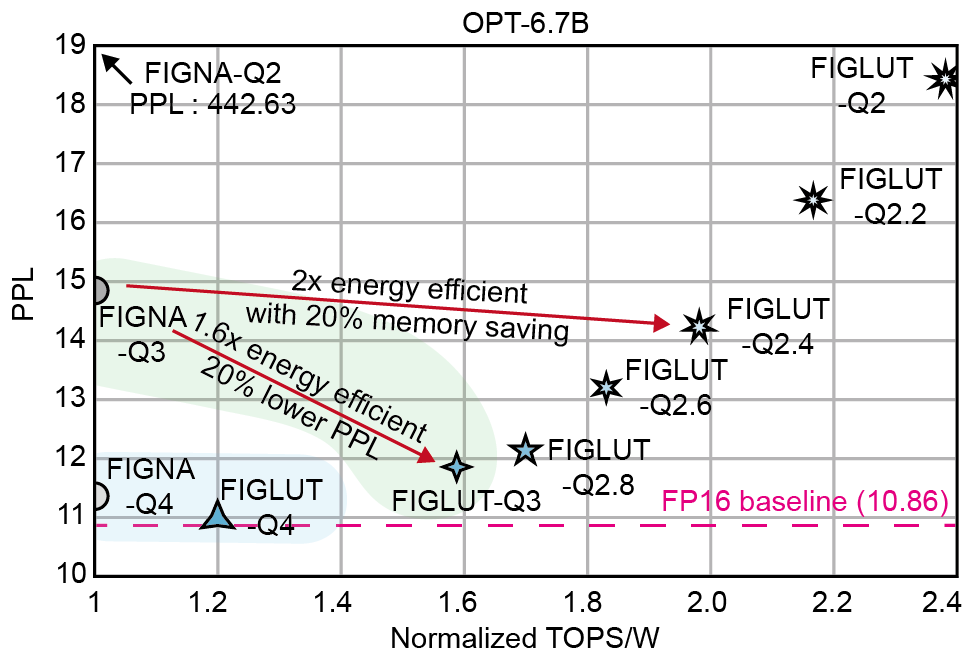}
  \caption{TOPS/W and perplexity score for various mixed-precision OPT-6.7B inference on FIGLUT.}
  \label{fig:mixedpre}
\end{figure}

Fig.~\ref{fig:mixedpre} shows the TOPS/W performance and perplexity results of FIGLUT across various mixed-precision configurations for the OPT-6.7B LLM.
We compare these results against the baseline FIGNA, evaluated using the OPTQ quantization method \cite{frantar2022optq} at 2, 3, and 4-bit precisions.
In contrast, FIGLUT, including mixed-precision performance, is assessed using ShiftAddLLM \cite{you2024shiftaddllm}.
Note that recent advancements in weight-only quantization techniques~\cite{frantar2022optq, lin2024awq, yuan2023rptq, chee2024quip, lee2023flexround, heo2023rethinking, kim2024memory, you2024shiftaddllm} have enabled sub-4-bit models to achieve performance levels comparable to the unquantized FP16 baseline, as indicated by the dotted lines in the figure.
Under the same 4-bit quantization, FIGLUT is $1.2\times$ more energy efficient with slightly better perplexity than FIGNA.
For Q3, FIGLUT achieves $1.6\times$ the energy efficiency of FIGNA and records a lower perplexity. 
Furthermore, at a similar energy efficiency level, FIGLUT-Q2.4 demonstrates $1.98\times$ the computational efficiency of FIGNA-Q3 while compressing the model size by 20\%.
For 2-bit weight precision, FIGLUT supports a non-uniform BCQ format, delivering much more stable perplexity performance compared to FIGNA, and improving energy efficiency by up to $2.4\times$.

\begin{table}[]
\centering
\caption{Comparison of hardware accelerator designs.}
\label{tab:gpu_comparison}
\begin{threeparttable}[]
\begin{tabular}{ccccc}
\hline
Hardware & \begin{tabular}[c]{@{}c@{}}Format\\ (Act.-Weight)\end{tabular} & \begin{tabular}[c]{@{}c@{}}Throughput\\ (TOPS)\end{tabular} & \begin{tabular}[c]{@{}c@{}}Power\\ (W)\end{tabular} & \begin{tabular}[c]{@{}c@{}}Energy Effi.\\ (TOPS/W)\end{tabular} \\ \hline
A100     & FP16-FP16                                                      & 40.27*                                                      & 192                                                 & 0.21*                                                            \\
A100     & FP16-Q4**                                                      & 1.85                                                      & 208                                                 & 0.01                                                            \\
H100     & FP16-FP16                                                      & 62.08*                                                      & 279                                                 & 0.22*                                                            \\
iFPU     & FP16-Q4                                                      & 0.14                                                        & 0.67                                                & 0.21                                                            \\
FIGNA    & FP16-Q4                                                      & 0.14                                                        & 0.41                                                & 0.33                                                            \\
FIGLUT  & FP16-Q4                                                      & 0.14                                                        & 0.29                                                & 0.47                                                            \\ \hline
\end{tabular}
\begin{tablenotes}[flushleft]
\item [*] The unit for GPUs is TFLOPS.
\item [**] LUT-GEMM kernel~\cite{park2022lut} is used.
\end{tablenotes}
\end{threeparttable}
\end{table}

Table~\ref{tab:gpu_comparison} shows the energy efficiency of various hardware accelerators.
Since GEMM operations dominate LLM inference workloads~\cite{park2022lut}, we focus on comparing GEMM performance.
For a fair comparison, we include state-of-the-art commercial accelerators, the A100 and H100 GPUs, additionally. 
As GPU architecture details are not publicly disclosed, we rely on empirical measurements for evaluation. 
GPU throughput is based on the computational workload and measured latency, while power consumption is obtained using the nvidia-smi query tool~\cite{park2022lut, tiwari2015reliability, ali2020evaluation}.
The target LLM model is OPT-6.7B, with a batch size of 32 for all evaluations to reflect realistic LLM usage.
To ensure fairness, we assume a 4-bit quantized weight format across all tests.
Note that, due to the absence of FP-INT units in GPUs, they rely on FP-FP units for computation with dequantization.
Although LUT-GEMM~\cite{park2022lut} proposes an efficient FP-INT kernel, it only supports a batch size of 1.
To standardize comparisons, we align all FP-Q4 accelerators to target the same maximum performance (TOPS). 

The H100 outperforms the A100 in terms of energy efficiency, primarily due to its advanced fabrication technology and increased memory bandwidth, which significantly boosts performance in memory-bound workloads like LLMs.
Note that the reported TFLOPS for GPUs are significantly lower than their theoretical peaks, primarily due to the small batch size.
In LLM inference infrastructure, batch sizes larger than 32 are rarely observed at a generation phase due to constraints in KV cache size and the optimization of batching systems for user requests~\cite{yu2022orca, kwon2023efficient}.
LUT-GEMM suffers from an even more pronounced performance drop as it operates with a batch size of 1 and relies on CUDA cores rather than Tensor Cores.
The iFPU, while designed to maximize efficiency with FP-INT units, demonstrates limited performance compared to the H100 due to the inherent constraints of bit-serial processing. 
FIGNA addresses these limitations, resulting in higher energy efficiency. 
Finally, FIGLUT further improves energy efficiency by overcoming the computational challenges of bit-serial processing through its LUT-based GEMM approach.
Given that FIGLUT uses a 28nm process technology while GPUs like the A100 and H100 leverage more advanced fabrication technology (7nm for A100, 4nm for H100), the efficiency of FIGLUT would be even more prominent if evaluated under comparable fabrication technologies.

\begin{table}[]
\centering
\caption{Perplexity comparison for weight-only quantization.}
\label{tab:bcq3_bcq4}
\begin{tabular}{lcccccc}
\hline
OPT     & \multicolumn{1}{c}{350M} & \multicolumn{1}{c}{1.3B} & \multicolumn{1}{c}{2.7B} & \multicolumn{1}{c}{6.7B} & \multicolumn{1}{c}{13B} & \multicolumn{1}{c}{30B} \\ \hline
FP16   & 22.00                    & 14.62                    & 12.47                    & 10.86                    & 10.13                   & 9.56                   \\
BCQ4~\cite{you2024shiftaddllm}   & 22.59                    & 15.11                    & 12.73                    & 11.08                    & 10.33                   & 9.70                   \\
BCQ3~\cite{you2024shiftaddllm}   & 28.72                    & 19.69                    & 15.28                    & 11.80                    & 10.70                   & 9.89                   \\ \hline
\end{tabular}
\end{table}

\section{Conclusion}
\textbf{Summary.}
This paper presents FIGLUT, an FP-INT accelerator that enhances the efficiency of bit-serial accelerators for weight-only quantized models using LUT-based operations.
We introduce a flip-flop based LUT (FFLUT) and replace the traditional MAC unit with a read-accumulate (RAC) unit for optimized performance.
Our analysis of LUT size and RAC configuration leads to an energy-efficient design that halves the LUT size through symmetry.
Experiments show FIGLUT outperforms state-of-the-art FP-INT accelerators in TOPS/W for sub-4-bit precision weights, significantly improving efficiency while maintaining flexibility in real-world systems.

\textbf{Limitations.}
As shown in Fig.~\ref{fig:energy_breakdown}, our design shows diminishing performance gains as weight bit precision increases, a limitation of the bit-serial architecture.
To achieve high performance with larger bit precisions, increasing $\mu$ would be necessary to fully leverage the computational benefits of LUT-based GEMM.
However, this results in much larger LUTs, increasing overhead in the LUT and its generator.
Nevertheless, Table~\ref{tab:bcq3_bcq4} indicates that state-of-the-art weight-only quantization techniques have enabled sub-4-bit quantization while maintaining accuracy comparable to the FP16 baseline.
We expect that continued advancements in weight-only quantization methods will further enhance the effectiveness of the proposed FIGLUT as bit precision is reduced.

\section{Acknowlegment}
This research was supported in part by the NAVER-Intel Co-Lab, conducted by POSTECH and reviewed by both NAVER and Intel.
This work was also supported by Institute of Information \& communications Technology Planning \& Evaluation (IITP) grant funded by the Korea government(MSIT) (No.RS-2023-00229849,MPU/Connectivity/TinyML SoC Solution for IoT Intelligence with Foundry-based eFLASH), and by National R\&D Program through the National Research Foundation of Korea(NRF) funded by Ministry of Science and ICT(RS-2023-00258227).



\bibliographystyle{IEEEtranS}
\bibliography{refs}

\end{document}